\documentclass[onecolumn,pra,showpacs,superscriptaddress]{revtex4}
\usepackage{amsmath}
\usepackage{graphicx}
\usepackage{dcolumn}
\usepackage{bm}

\begin{document}
\title{Conditional sign flip via teleportation}
\author{Gian Luca Giorgi}
\author{Ferdinando  de Pasquale}
\affiliation{INFM Center for Statistical Mechanics and Complexity}
\affiliation{Dipartimento di Fisica, Universit\`{a} di Roma La
Sapienza, P. A. Moro 2, 00185 Rome, Italy}
\email{gianluca.giorgi@roma1.infn.it}
\author{S. Paganelli}
\affiliation{Dipartimento di Fisica, Universit\`{a} di Roma La
Sapienza, P. A. Moro 2, 00185 Rome, Italy}
\affiliation{Dipartimento di Fisica, Universit\`{a} di Bologna,
Via Irnerio 46, I-40126, Bologna, Italy}


\begin{abstract}
We present a model to realize a probabilistic conditional sign
flip gate using only linear optics. The gate operates in the space
of number state qubits and is obtained by a nonconventional use of
the teleportation protocol. Both a destructive and a
nondestructive version of the gate are presented. In the former
case an Hadamard gate on the control qubit is combined with a
projective teleportation scheme mixing control and target. The
success probability is 1/2. In the latter case we need a quantum
encoder realized via the interaction of the control qubit with an
ancillary state composed of two maximally entangled photons. The
success probability is 1/4.
\end{abstract}
\maketitle
\section{Introduction}

Single photon qubits are a promising tool for quantum computation
\cite {nielsen}. The great advantage with respect to the others
physical implementation \cite{nakamura,loss,vandersypen} is
represented by the fact that photonic systems can be easily
transferred from one place to another in the space and moreover
the weak interaction with the environment makes the decoherence
not so dangerous. These features permit secure transmission of
information over long distances \cite{bb84,ekert}. On the other
hand, the robustness of photons with respect to interactions
creates a serious obstacle to the realization of conditional gates
essential for quantum computation \cite{divincenzo} due to the
large amount of resources required to create nonlinear coupling
between qubits. Despite these considerations, Knill, Laflamme and
Milburn (KLM) showed that quantum computation can be realized
using only linear optics \cite{KLM}. This is done exploiting the
nonlinearity induced by a measurement process. Probabilistic
conditional gates are obtained using single photon sources, single
photon detectors, ancilla photons and postselection measurements.
More recently, Nielsen showed that any probabilistic gate based on
linear optics is sufficient for the implementation of a quantum
computer \cite{nielsen2}.

There is also some experimental realization of gates using only
linear optics: Controlled-Not gate \cite{obrien,pittman,gasparoni}
and Nonlinear sign shift \cite{sanaka} have been recently
reported.

On the other hand, it's generally accepted that the teleportation
protocol \cite{bennett} represents a fundamental resource for
quantum computation, as already shown by Gottesman and Chuang
\cite{gottesman}.

Here we propose a model for a conditional sign flip gate based on
photon number qubits, in agreement of most of features of KLM
protocol, based on a nonconventional use of teleportation process.

In section \ref{II} we show how a destructive C-sign gate
\cite{pittman2} can be implemented starting from an Hadamard gate
on the control qubit and a projective teleportation mixing control
and target. How to obtain a non destructive gate is the subject of
section \ref{III}, while section \ref{IV} will be devoted to
conclusions.

\section{Destructive C-sign Flip gate\label{II}}

A conditional sign flip gate is a two-qubit gate: the target qubit
experiences a sign change between its components $\left|
0\right\rangle $ and $\left| 1\right\rangle $\ if and only if the
control qubit is in the logic state $\left| 1\right\rangle $. In
the basis $\left\{ \left| 00\right\rangle ,\left| 01\right\rangle
,\left| 10\right\rangle ,\left| 11\right\rangle \right\} $ the
unitary operator representing the gate is $U=\left| 0\right\rangle
\left\langle 0\right| ^{\left( 1\right) }\otimes I^{\left(
2\right) }+\left| 1\right\rangle \left\langle 1\right| ^{\left(
1\right) }\otimes \sigma _{z}^{\left( 2\right) }$ ( $I$ and
$\sigma _{z}$ are respectively the identity operator and one of
Pauli matrices) and has the following matrix representation:
\begin{equation}
U=\left(
\begin{array}{cccc}
1 & 0 & 0 & 0 \\
0 & 1 & 0 & 0 \\
0 & 0 & 1 & 0 \\
0 & 0 & 0 & -1
\end{array}
\right)  \label{U}
\end{equation}
On the other hand, the teleportation can be briefly described as follows. A quantum state $%
\left| \alpha _{1}\right\rangle =a\left| 0_{1}\right\rangle
+b\left| 1_{1}\right\rangle$ is combined with a two qubit
maximally entangled  Bell state $\left| \Psi _{23} \right\rangle$.
A Bell measurement, performed on the qubits $1$ and $2$, causes
the transfer on the third qubit of the superposition initially
encoded on the first one, except for a unitary transformation
determined by the result of the Bell measurement. From a formal
point of view, the teleportation is represented by a base change
in the combined Hilbert space $\mathcal{H}_{1}\otimes
\mathcal{H}_{2}\otimes \mathcal{H}_{3}$, plus a measurement.
Usually the state $\left| \Psi _{23} \right\rangle$ is considered
as fixed, but this is not a necessary prescription. In a more
complete description, the global input state is written in terms
of all possible Bell states, each of them with a probability
amplitude $u_i$ where $i=0,z,x,y$ (the choice of symbols will
appear clear in what follows), that we can use to perform the
process: recalling that the Bell states are $\left| \Phi ^{\pm }\right\rangle =1/%
\sqrt{2}\left( \left| 00\right\rangle \pm \left| 11\right\rangle
\right) $ and $\left| \Psi ^{\pm }\right\rangle =1/\sqrt{2}\left(
\left| 10\right\rangle \pm \left| 01\right\rangle \right) $) we
have
\begin{equation}
\left| \Phi \right\rangle =\left| \alpha \right\rangle _{1}\left(
u_{0}\left| \Psi _{23}^{+}\right\rangle +u_{z}\left| \Psi
_{23}^{-}\right\rangle +u_{x}\left| \Phi _{23}^{+}\right\rangle
+u_{y}\left| \Phi _{23}^{-}\right\rangle \right)
\end{equation}
After the base change we obtain a new expression in terms of Bell
states on $1$ and $2$:
\begin{equation}
\left| \Phi \right\rangle =\sum_{i}\left( \left| \Psi
_{12}^{+}\right\rangle u_{i}a_{0i}\sigma _{i}\left| \alpha
_{3}\right\rangle +\left| \Psi _{12}^{-}\right\rangle
u_{i}a_{zi}\sigma _{z}\sigma _{i}\left| \alpha _{3}\right\rangle
+\left| \Phi _{12}^{+}\right\rangle u_{i}a_{xi}\sigma _{x}\sigma
_{i}\left| \alpha _{3}\right\rangle +\left| \Phi
_{12}^{-}\right\rangle u_{i}a_{yi}\sigma _{y}\sigma _{i}\left|
\alpha _{3}\right\rangle \right)
\end{equation}
having introduced the Pauli matrices acting on the third qubit
and
\begin{equation*}
a_{ij}=\left(
\begin{array}{cccc}
1 & -1 & 1 & i \\
1 & -1 & -i & -1 \\
1 & -i & 1 & 1 \\
-i & 1 & 1 & 1
\end{array}
\right)
\end{equation*}
If a measurement is done by projection, e.g. on the  the singlet
state $\left| \Phi _{12}^{-}\right\rangle$, we obtain a different
state of the third qubit according to the $u_i$ selected. This
result shows that teleportation acts as a controlled gate: the
teleported state experiences a unitary transformation determined
by the Bell state used as an input. In the most general case both
C-NOT and C-sign are contemplated respectively when $u_0$ and
$u_x$ or $u_0$ and $u_z$ are nonvanishing and by the establishment
of a connection between the logic value of a qubit used as control
and the suitable pair of Bell states $\left|
\Psi_{23}\right\rangle$ selected. In particular, we found a simple
model where this behavior emerges giving rise to a C-sign flip
gate.

Let us explain our proposal. As requested in \cite{KLM} each qubit
is realized on two spatial modes \cite{gianluca,yamamoto}: the
presence of the photon in the first (second) rail corresponds to
the logic state $\left| 1\right\rangle $ ($\left| 0\right\rangle
$). For the sake of clarity we shall utilize the second
quantization language, using occupation numbers instead of logic
values, writing $\left| 01\right\rangle $ for $\left|
0\right\rangle $\ and $\left| 10\right\rangle $ for $\left|
1\right\rangle $.

The rails of the control qubit are the input arms of a 50\% beam splitter ($%
BS_{1}$) that acts as an Hadamard gate:
\begin{equation}
H=\frac{1}{\sqrt{2}}\left(
\begin{array}{cc}
1 & 1 \\
-1 & 1
\end{array}
\right)
\end{equation}
Then, if the input photon is in the state $\left| 01\right\rangle
$ the
output state is an entangled singlet state, while if it is in the state $%
\left| 10\right\rangle $ we deal with a triplet one on the output arms.

\begin{figure}
  \includegraphics[width=3.9075in]{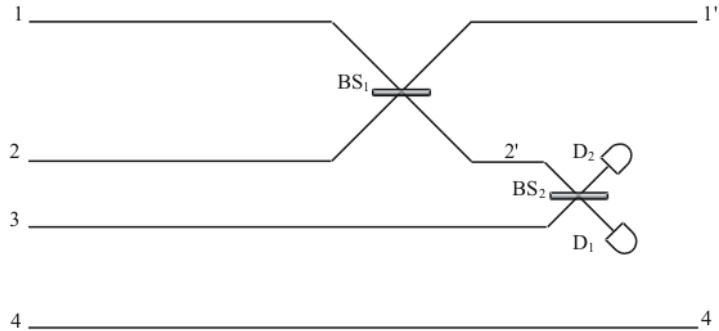}
  \caption{Destructive conditional sign flip gate: the modes $1$ and $2$ correspond to the
control qubit, while the modes $3$ and $4$ correspond to the
target qubit. The beam splitter $BS_{1}$ acts as an Hadamard gate on the control qubit and $%
BS_{2}$ is used to perform quantum teleportation.}\label{fig1}
\end{figure}

The entangled states created are used to perform teleportation. We refer to
the experimental realization of vacuum-one photon qubit teleportation \cite
{lombardi}. One of spatial modes outgoing from $BS_{1}$ is mixed on a second
50\% beam splitter ($BS_{2}$) with one of spatial modes of the target qubit.
With reference to  figure \ref{fig1}, we denote with $1$ and $2$ the modes
associated to the control qubit, with $1^{\prime }$ and $2^{\prime }$ the
output modes of $BS_{1}$ and with $3$ and $4$ the modes corresponding to the
target qubit, while the output modes of $BS_{2}$ will be labelled with $5$
and $6$.
Let us consider first the case that the control qubit is in the state $%
\left| 1_{1}0_{2}\right\rangle $. Due to the action of the Hadamard gate the
state after the photon has impinged $BS_{1}$ is $1/\sqrt{2}\left( \left|
0_{1^{\prime }}1_{2^{\prime }}\right\rangle +\left| 1_{1^{\prime
}}0_{2^{\prime }}\right\rangle \right) $. This is a triplet entangled state
realized over the output spatial modes of $BS_{1}$.

Being the target qubit in an arbitrary superposition $\alpha
\left| 0_{3}1_{4}\right\rangle +\beta \left|
1_{3}0_{4}\right\rangle $ the whole state is
\begin{equation}
\left| \Psi \right\rangle =\frac{1}{\sqrt{2}}\left( \alpha \left|
0_{1^{\prime }}1_{2^{\prime }}0_{3}1_{4}\right\rangle +\beta \left|
0_{1^{\prime }}1_{2^{\prime }}1_{3}0_{4}\right\rangle +\alpha \left|
1_{1^{\prime }}0_{2^{\prime }}0_{3}1_{4}\right\rangle +\beta \left|
1_{1^{\prime }}0_{2^{\prime }}1_{3}0_{4}\right\rangle \right)
\end{equation}

The portion of this state corresponding to the spatial modes $2^{\prime }$
and $3$ is conveniently rewritten in terms of Bell states $\left| \Phi ^{\pm
}\right\rangle =1/\sqrt{2}\left( \left| 00\right\rangle \pm \left|
11\right\rangle \right) $\ and $\left| \Psi ^{\pm }\right\rangle =1/\sqrt{2}%
\left( \left| 10\right\rangle \pm \left| 01\right\rangle \right) $. After
this substitution we have
\begin{eqnarray}
\left| \Psi \right\rangle &=&\frac{1}{2}[\left| \Psi _{2^{\prime
}3}^{+}\right\rangle \left( \alpha \left| 0_{1^{\prime }}1_{4}\right\rangle
+\beta \left| 1_{1^{\prime }}0_{4}\right\rangle \right) +\left| \Psi
_{2^{\prime }3}^{-}\right\rangle \left( \alpha \left| 0_{1^{\prime
}}1_{4}\right\rangle -\beta \left| 1_{1^{\prime }}0_{4}\right\rangle \right)
+  \nonumber \\
&&\left| \Phi _{2^{\prime }3}^{+}\right\rangle \left( \alpha \left|
1_{1^{\prime }}1_{4}\right\rangle +\beta \left| 0_{1^{\prime
}}0_{4}\right\rangle \right) +\left| \Phi _{2^{\prime }3}^{-}\right\rangle
\left( \alpha \left| 1_{1^{\prime }}1_{4}\right\rangle -\beta \left|
0_{1^{\prime }}0_{4}\right\rangle \right) ]  \label{2}
\end{eqnarray}
Our idea is to perform a projective measurement over the modes
$2^{\prime }$ and $3$\ by selecting only those events
corresponding to the state $\left| \Psi _{2^{\prime
}3}^{-}\right\rangle $\ as result. The measurement is
performed using the modes $2^{\prime }$ and $3$ as the input arms of $BS_{2}$%
. The state $\left| \Psi _{2^{\prime }3}^{-}\right\rangle $ corresponds to
the detection of one and only one photon on the detector $D_{1}$ (see figure
1) and to the absence of counts on the second detector $D_{2}$.

As a result, the state emerging on the spatial modes $1^{\prime }$ and $4$
is $\alpha \left| 0_{1^{\prime }}1_{4}\right\rangle -\beta \left|
1_{1^{\prime }}0_{4}\right\rangle $. We observe that an entanglement
swapping has been realized together with a sign flip with respect to the
incoming target state.

Next, we study the situation corresponding to a control qubit in the $\left|
0_{1}1_{2}\right\rangle $. In such a situation the Hadamard gate creates a
singlet entangled state on the output modes of $BS_{1}$: $1/\sqrt{2}\left(
\left| 0_{1^{\prime }}1_{2^{\prime }}\right\rangle -\left| 1_{1^{\prime
}}0_{2^{\prime }}\right\rangle \right) $. Then Eq. \ref{2} has to be
opportunely modified. Limiting our interest to the term associated with the
singlet as output result, now we have $\left| \Psi _{2^{\prime
}3}^{-}\right\rangle \left( \alpha \left| 0_{1^{\prime }}1_{4}\right\rangle
+\beta \left| 1_{1^{\prime }}0_{4}\right\rangle \right) $. Thus, we observe
again an entanglement swapping, but the difference with the former situation
is that no sign flip arises from the process.

The previous results can be synthesized stating that the target
qubit, initially encoded using the modes $3$ and $4$, is
transferred on $1^{\prime } $ and $4$ with a sign change
conditional to the logic state of the control qubit, as required
from the definition of the C-sign gate. The gate is deterministic:
it does not work with a success probability equal to 1, but we
know whether it works correctly. In our case the probability is
1/4, determined by the postselection procedure selecting one of
four Bell states, and it can increased up to 1/2 accepting single
counts on $D_{2}$, with an adjunctive single qubit rotation.

Unluckily, the control qubit is destroyed by the projection and
the gate above illustrated is not complete. To make the scheme
useful for quantum computation a method to restore the control
state has to be introduced.

\section{Nondestructive gate\label{III}}

To overcome the previous obstacle we use the technique of quantum encoding.
From the ``no cloning theorem'' \cite{wootters}\ we learn that a physical
machine able to copy an arbitrary quantum state in a blank state cannot be
realized. However, the theorem does not exclude the possibility of copying
two selected orthogonal states and this is the working principle of a
quantum encoder. Roughly speaking, the conversion $\left( \alpha \left|
0\right\rangle +\beta \left| 1\right\rangle \right) \rightarrow \left(
\alpha \left| 0\right\rangle +\beta \left| 1\right\rangle \right) \otimes
\left( \alpha \left| 0\right\rangle +\beta \left| 1\right\rangle \right) $
is forbidden while $\left( \alpha \left| 0\right\rangle +\beta \left|
1\right\rangle \right) \rightarrow \left( \alpha \left| 0\right\rangle
\otimes \left| 0\right\rangle +\beta \left| 1\right\rangle \otimes \left|
1\right\rangle \right) $ is \ (at least in a probabilistic way) allowed
leaving $\alpha $ and $\beta $\ out of consideration.

A quantum encoder operating on polarization qubits is described in
\cite {pittman2,pittman3}. It applies also in our case due to the
existence of converters from polarization to dual rail and vice
versa that are easily realizable using a polarizing beam splitter
and a $\lambda /2$ waveplate.
\begin{figure}
  \includegraphics[width=3.9075in]{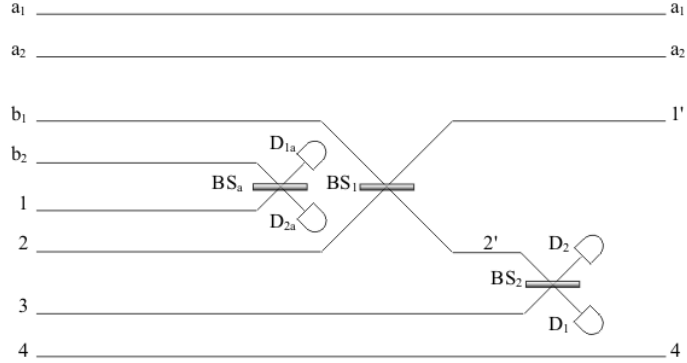}
  \caption{Nondestructive conditional sign flip gate: the modes $a_{1}$, $a_{2}$, $b_{1}$ and $%
b_{2}$ represent the quantum encoder, control and target qubit are
yet implemented respectively on the modes $1$ and $2$ and $3$ and
$4$ . The auxiliary beam splitter $BS_{a}$ and the auxiliary
detectors  $D_{a_{1}}$ and $D_{a_{2}}$ are used to ``double'' the
control qubit in an entangled state on $a_{1}$, $a_{2}$, $b_{1}$
and $2$. $BS_{1}$ and $BS_{2}$ perform the conditional gate, as
illustrated in \ref{fig2}, and the output is represented by the
control qubit on the modes $a_{1}$and $a_{2}$ and the (modified by
the gate) target qubit on the modes $1^{\prime }$and
$4$.}\label{fig2}
\end{figure}

On the other hand, we will show that a quantum encoder working
only with photon number qubits is feasible using non polarizing
beam splitters. The scheme is depicted in figure \ref{fig2}. The
control qubit $\left( \alpha _{1}\left| 01\right\rangle +\alpha
_{2}\left| 10\right\rangle \right) $ we want to copy
is defined on the modes $1$ and $2$, while modes $a_{1}$, $a_{2}$, $b_{1}$, $%
b_{2\text{ }}$correspond to two ancilla qubits previously prepared in an
maximally entangled state $1/\sqrt{2}\left( \left|
0_{a_{1}}1_{a_{2}}0_{b_{1}}1_{b_{2}}\right\rangle -\left|
1_{a_{1}}0_{a_{2}}1_{b_{1}}0_{b_{2}}\right\rangle \right) $. The modes $b_{2%
\text{ }}$and $1$ are mixed on a beam splitter ($BS_{a}$) and a projective
measurement analogous to that one described in section II takes place
selecting only the singlet state $\left| \Psi _{b_{2}1}^{-}\right\rangle =1/%
\sqrt{2}\left( \left| 0_{b_{2}}1_{1}\right\rangle -\left|
1_{b_{2}}0_{1}\right\rangle \right) $. The projection is performed
measuring one and only one photon on $D_{a_{1}}$ and zero photons
on $D_{a_{2}}$.\ As a result of \ the projection, it remains
$1/\sqrt{2}\left( \alpha _{1}\left|
0_{a_{1}}1_{a_{2}}0_{b_{1}}1_{2}\right\rangle +\alpha _{2}\left|
1_{a_{1}}0_{a_{2}}1_{b_{1}}0_{2}\right\rangle \right) $. Thus, we
have
realized the quantum encoding operation, apart from a swapping from mode $1$%
\ to $b_{1}$. This gate is probabilistic being conditioned from
the output of the Bell measurement. The success probability is 1/4
and again it reaches 1/2 if also $\left| \Psi
_{b_{2}1}^{+}\right\rangle =1/\sqrt{2}\left( \left|
0_{b_{2}}1_{1}\right\rangle +\left| 1_{b_{2}}0_{1}\right\rangle
\right) $ is accepted via a classically feed-forwarded one qubit
rotation. Notice that a qubit can be encoded also on a string of
$n$ qubits simply using a generalized maximally entangled state
$1/\sqrt{2}\left( \left| 0101.....01\right\rangle -\left|
1010.....10\right\rangle \right) $ and performing the projection
measurement mixing one of the $2n$ modes with one mode of the
incoming qubit.

Let us return to our main problem. We want to build a gate that transforms a
two qubit state, defined on four spatial modes, in accordance with the
operator $U$ introduced in Eq. \ref{U}:
\begin{equation}
U\left( \alpha _{1}\left| 0_{1}1_{2}\right\rangle +\alpha _{2}\left|
1_{1}0_{2}\right\rangle \right) \left( \alpha _{3}\left|
0_{3}1_{4}\right\rangle +\alpha _{4}\left| 1_{3}0_{4}\right\rangle \right)
=\alpha _{1}\alpha _{3}\left| 0_{1}1_{2}0_{3}1_{4}\right\rangle +\alpha
_{1}\alpha _{4}\left| 0_{1}1_{2}1_{3}0_{4}\right\rangle +\alpha _{2}\alpha
_{3}\left| 1_{1}0_{2}0_{3}1_{4}\right\rangle -\alpha _{2}\alpha _{4}\left|
1_{1}0_{2}1_{3}0_{4}\right\rangle
\end{equation}

The control state is doubled via the quantum encoder above
introduced and, under the probabilistic condition relied to the
postselection process, we deal with the initialized three qubit
state
\begin{equation}
\left| \Psi \right\rangle =\left( \alpha _{1}\left|
0_{a_{1}}1_{a_{2}}0_{b_{1}}1_{2}\right\rangle +\alpha _{2}\left|
1_{a_{1}}0_{a_{2}}1_{b_{1}}0_{2}\right\rangle \right) \left(
\alpha _{3}\left| 0_{3}1_{4}\right\rangle +\alpha _{4}\left|
1_{3}0_{4}\right\rangle \right)
\end{equation}

The procedure described in the previous section can now start: the
modes $b_{1}$ and $2$ are rearranged in $1^{\prime }$ and
$2^{\prime }$ via the $BS_{1}$, the modes $2^{\prime }$ and $3$
are mixed on $BS_{2}$, the postselection measurement on $\left|
\Psi _{2^{\prime }3}^{-}\right\rangle $ is performed, and as a
result of the complete set of operations we find that $U$ creates
the state
\[
\alpha _{1}\alpha _{3}\left| 0_{a_{1}}1_{a_{2}}0_{1^{\prime
}}1_{4}\right\rangle +\alpha _{1}\alpha _{4}\left|
1_{a_{1}}0_{a_{2}}1_{1^{\prime }}0_{4}\right\rangle +\alpha _{2}\alpha
_{3}\left| 1_{a_{1}}0_{a_{2}}0_{1^{\prime }}1_{4}\right\rangle -\alpha
_{2}\alpha _{4}\left| 0_{a_{1}}1_{a_{2}}1_{1^{\prime }}0_{4}\right\rangle
\]
in perfect agreement with the definition of the C-sign flip gate.
Furthermore, the scheme realizes a teleported gate, as outlined in
\cite {gottesman}.

Due to the nondeterministic nature of the destructive gate and the
quantum encoder, the nondestructive C-sign flip can reach 1/4 as
overall efficiency.

\section{Conclusions\label{IV}}

We have proposed a method to realize a probabilistic C-sign flip
gate for number state qubits based only on few linear optics
elements, specifically three balanced beam splitters, one source
of entangled photons for auxiliary states, two single photon
sources for target and control qubits, photodetectors and
postselection measurements. All these elements are contained in
the KLM scheme, for which our model seems to be tailored. In the
original proposal contained in \cite{KLM} the C-sign gate was
achieved via two Nonlinear sign shift combined with two beam
splitters. The network created in such a scheme was very
intricate, and the simplification arising from the idea previously
illustrated is remarkable. Furthermore, the maximum success
probability of the gate is the same reported in the KLM work. To
achieve the gate, a four fold coincidences measurement is
required, fully available with the present technology. This
scheme, being based on manipulations of number states, could be
extended to solid state devices, where the degenerate ground state
is used both for transferring information and performing the
unitary rotation associated to a beam splitter \cite{ferdy}.

\end{document}